\documentclass[10pt,lcy,amsmath,amssymb,twocolumn]{revtex4}
\newcommand{\dfr}[2]{\frac {\displaystyle #1}{\displaystyle #2}}
\usepackage{graphicx}

\begin{document}
\title{Quantum critical dynamics of the boson system in the Ginsburg-Landau model}
\author{Vasin M.G.}
\affiliation{Physical-Technical Institute, Ural Branch of RAS, 426000 Izhevsk, Russia}
\affiliation{Institute for High Pressure Physics of RAS, 142190, Moscow, Russia}

\begin{abstract}
The quantum critical dynamics of the quantum phase transitions is considered.
In the framework of the unified theory, based on the Keldysh technique, we consider the crossover from the classical to the quantum description of the boson many-particle system dynamics close to the second order quantum phase transition.
It is shown that in this case the upper critical space dimension of this model is $d_c^{+}=2$, therefore the quantum critical dynamics approach is useful  in case of $d<2$.  In the one-dimension system the phase coherence time does diverge at the quantum critical point, $g_c$, and has the form of $\tau \propto -\ln |g-g_c|/(g-g_c) $, the correlation radius diverges as $r_c\propto |g-g_c|^{-\nu } (\nu = 0.6)$.
\end{abstract}

\maketitle

\section{Introduction}

We consider the dissipative critical dynamics of the quantum phase transitions (QPT) taking place in the system of the coupled enharmonic oscillators with one-component order parameter ($n=1$) corresponding, for example, to the Ising magnet \cite{Sachdev}.

It is well known that at $T=0$, in the regime of quantum fluctuations (zero-point fluctuations), the ordering phase transition is possible in these systems \cite{Kagan}.
In addition it is believed that the critical exponents of this phase transition are determined with help of the simple rule: the exponents of the phase transition in a $d$-dimension system at $T=0$ are the same as at  $T\neq 0$ but in the system with greater per unit dimension: $d_{eff}=d+z=d+1$ \cite{Pankov}. Hence one can conclude that the upper critical space dimension of the considered system is $d^+_{cr}=3$. Let us call it quantum mechanical (QM) approach. However, when describing the dynamics of the statistical ensemble of the coupled oscillators ($N\to \infty$) one needs to take into account the dissipation effect \cite{Weiss}. In case of $\hbar \omega \gg kT$ this leads to the change of the critical exponents and the universality class of the phase transition in the one-dimension system: $z\thickapprox 2$, $d^+_{cr}=2$ \cite{HH}.

In this paper we describe the critical dynamics of the Ising magnet system close to QPT using the Keldysh technique \cite{K}. This approach is developed for description of the dynamics of non-equilibrium quantum systems. Therefore, we suppose it will allow us to describe the crossover from the critical dynamics to the quantum critical dynamics (QCD) using the uniform technique. We also believe that it help us to outline the borders of applicability of the QM and QCD approaches to the QPT description.

\section{Crossover from the critical dynamics to the quantum critical dynamics in the Keldysh technique}
Let us consider the quantum critical dynamics of the Ginsburg-Landau  model in terms of the Keldysh technique.
The Lagrangian of this model has the following form:
\begin{gather}
    \mathcal{L}\thickapprox(\vec\partial \phi )^2+\mu(g)\phi^2+v(g)\phi^4.
\label{I}
\end{gather}
where $\phi $ is the scalar order parameter field, which obeys to the bose statistics.
We suppose $\mu $ and $v$ to depend on some external parameter $g$, that controls the system state.

It is convenient to describe the non-equilibrium dynamics of quantum systems in terms of the Keldysh technique.
Since we assume the uniform description of both quantum ($T\to 0$) and classical ($T\gg 0$) systems, we consider the system interacting with a heat bath at the temperature $T$.

According to the Keldysh approach to the description of non-equilibrium dynamics of the system one should write the generating functional in the form of
\begin{gather*}
    W=\int \mathfrak{D}\vec\phi \exp\left\{i\int d^{d+1}x \mathcal{L}(\phi_{cl},\,\phi_q;\,g_{cl},\,g_q)\right\},
\end{gather*}
where $\vec \phi=\{\phi_q,\,\phi_{cl}\}$, $\phi_{cl}$ and $\phi_q $ are the ``classical'' and ``quantum'' parts of the order parameter accordingly, $g_{cl}$ and $g_q$ are the sources of these fields, and $\mathcal{L}$ is the fields lagrangian density.
Below it will be more convenient to move from the Minkowski space to  the Euclidean one by the Wick rotation, $t= -ix_4 $. Then
\begin{gather*}
    W=\int \mathfrak{D}\vec\phi \exp\left\{-\int d^{d}kd\omega \mathcal{L}(\phi_{cl},\,\phi_q;\,g_{cl},\,g_q)\right\},
\end{gather*}
Note, that in this case every contact of the system with any environment, including external noise, is described as the interaction with the heat bath, while the ``internal (quantum) noise'' is implied in the description directly in the description.
In this case according to \cite{K} one can write the Keldysh Lagrangian in the form of:
\begin{gather*}
    \mathcal{L}=\mathcal{L}_{free}+\mathcal{L}_{int}+\mathcal{L}_{noise},
\end{gather*}
where
\begin{gather*}
   \mathcal{L}_{free}=\phi_q\left( \varepsilon_k-i\gamma \omega\right)\phi_{cl}+\phi_{cl}\left( \varepsilon_k+i\gamma \omega\right)\phi_q,\\[10pt]
   \mathcal{L}_{int}= -\frac{}{}U(\phi_{cl}+\phi_q,\,g_{cl}+g_q)+U(\phi_{cl}-\phi_q,\,g_{cl}-g_q),\\
   \mathcal{L}_{noise}=\phi_q \left(2\gamma \omega \coth \dfr{\omega }{T}\right)\phi_q,
\end{gather*}
$\varepsilon_k=k^2+\mu(g)$, and $U(\phi)$ is the interaction part.

According to the Keldysh approach to the description of non-equilibrium dynamics one can write an expression for the Retard, Advanced and the Keldysh parts of the Green function (matrix) in the form of:
\begin{gather*}
    G^K=G^R\circ F-F\circ G^A ,
\end{gather*}
where $F$ is the Hermitian matrix ($F=F^{\dag}$), and the circular multiplication sign implies integration over the intermediate time (matrix multiplication) \cite{K}.
One can check that
\begin{gather*}
    [G^{-1}]^K=[G^R]^{-1}\circ F-F\circ [G^A]^{-1} .
\end{gather*}
After the Wigner transform (WT) in the frequency representation we come to
\begin{gather*}
    G^K=f(\omega )(G^R-G^A) ,\\
    [G^{-1}]^K=f(\omega)\left([G^R]^{-1}-[G^A]^{-1}\right),
\end{gather*}
where $f(\omega )$ is the distribution function. For a boson system in thermal equilibrium $f=-i\coth (\omega/T)$,
where $T$ is the temperature of the heat bath \cite{K}. This is the FDT, which, as it is shown later, takes a different form in the classical and quantum limits.

\begin{figure}[h]
\centering
   \includegraphics[scale=1.5]{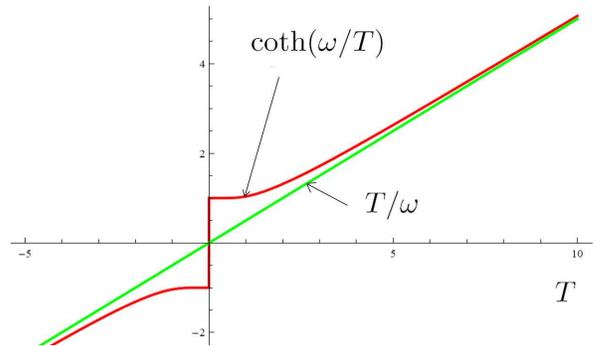}
   \caption{The red line is the graphic representation of $\coth (\omega/T)$ versus $T$ function (with $\omega =4$), the green line is the $T/\omega $ function. At high temperatures these graphics coincide, which corresponds to the critical dynamics. However, $\coth (\omega/T)\to \mbox{sign}(\omega )$ close to $T=0$, where the system is described by the quantum critical dynamics.}
   \label{f1}
\end{figure}

If we consider the system with dissipation, then
\begin{gather}
    [G^R]^{-1}=\varepsilon_k +i\gamma\omega ,\quad [G^A]^{-1}=\varepsilon_k -i\gamma\omega ,\\
    [G^{-1}]^K=2\gamma\omega \coth (\omega/T),
\label{a3}
\end{gather}
where $\gamma $ is the kinetic coefficient.
In the quantum case $T\ll \omega $ (see Fig.\,\ref{f1})
\begin{gather*}
    \coth (\omega/T)\to \mbox{sign}(\omega ) \quad \Rightarrow \quad [G^{-1}]^K=2\gamma |\omega |.
\end{gather*}
The FDT has the following form: $ G^K=i\,\mbox{sign}(\omega )(G^R-G^A)$.
In the classical case $T\gg \omega$ (see Fig.\,\ref{f1})
\begin{gather*}
    \coth (\alpha \omega)\to \dfr T{\omega } \quad \Rightarrow \quad [G^{-1}]^K=2\gamma T,
\end{gather*}
and the system satisfies the usual classical form of FDT: $G^K=T(G^R-G^A)/i\omega $.

Below we will concentrate on the quantum limit ($\omega \gg T\approx 0$), when $\coth(\omega /T)\to \mbox{sign} (\omega )$,
the temperature is not essential in the FDT, $\mathcal{L}_{noise}=\phi_q \left(2\gamma |\omega |\right)\phi_q$,
and the Keldysh Green function has the following form:
\begin{gather*}
    G^K(\omega )=\dfr {2\gamma |\omega |}{\varepsilon_k^2 +\gamma^2\omega^2}.
\end{gather*}
Note, that in the case of $k\to 0$ $G^K(\omega )={2}/{\gamma |\omega |}$.
This is the so called $1/f$-noise, whose intensity does not depend on the temperature but it is equal to $\hbar$.
One can infer that the presence of $1/f$-noise is a natural property of the cold many-body bose system, which follows from the quantum character of dynamics in $T=0$.

\section{Quantum critical dynamics of the $d=2-\varepsilon $ Ginsburg--Landau model}
We suppose that the system is close to the second order phase transition, when the interaction part of the action can be written as
\begin{gather}
U\approx\mu(g) \phi^2+v(g_c)\phi^4,
\end{gather}
where $\mu(g)=(g-g_c) \to 0$ close to the phase transition point, $g_c$.

Below we will consider the quantum limit $T\to 0$ of the critical dynamics of this system in the $d=2-\varepsilon$ space dimension close to the second order critical point.
The critical dynamics rests on the hypothesis of dynamic scaling, according to which the action should be invariant with respect to the scale transformations which conformly expand the space and time coordinates ($\omega \propto k^{d_{\omega }}$). In this case the summarized dimension, $D=d+d_{\omega }$ ($d_{\omega}=z$ is the dynamic exponent), has the same role as the conventional (momentum) dimension, $d_k$, in the static case. The canonical dimensions of the fields and the model parameters are determined from the condition of dimensionless action. The corresponding summarized canonical dimensions, $D[F]$, of any values, $F$, are determined as:
$$
D[F]=d[F]+z\cdot d_{\omega }[F],
$$
where $d_{\omega }[F]$ is the frequency dimension \cite{Vas,Pat}.
The canonical dimensions of the values of our theory are given in the table:
\begin{center}
\begin{tabular}{|c|c|c|c|c|c|c|c|}
  \hline
  $F$ & $k$ & $\omega$ & $\phi_{\rm cl}$ & $\phi_{\rm q}$ & $v$ & $\gamma $ & $\mu $  \\ \hline
  $d[F]$ & 1 & 0 & $-2+\varepsilon/2$ & $-2+\varepsilon/2$ & $2+\varepsilon$ & $2$ & $2$  \\ \hline
  $d_{\omega }[F]$ & $0$ & $1$ & $-1/2$ & $-1/2$ & $-1$ & $-1$ & $0$  \\ \hline
  $D[F]$ & $1$ & $z=2$ & $-3+\varepsilon/2$ & $-3+\varepsilon/2$ & $\varepsilon$ & $0$ & $2$ \\
  \hline
\end{tabular}
\end{center}

\begin{figure}[h]
\centering
   \includegraphics[scale=0.9]{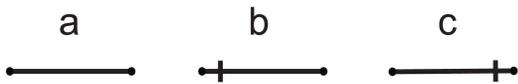}
   \caption{The graphic representation of the Keldish, $G^K$ (a), advanced, $G^A$ (b), and retarded, $G^R$ (c), Green functions of the theory.}
   \label{f2}
\end{figure}
\begin{figure}[h]
\centering
   \includegraphics[scale=0.6]{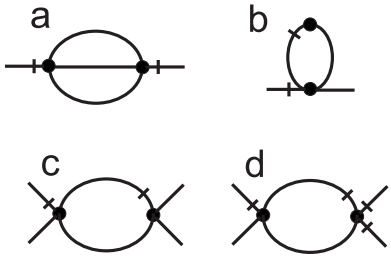}
   \caption{ The graph representation of the contributions to the renormalization of the theory's vertexes.}
   \label{f3}
\end{figure}

The renormalization procedure is carried out with the standard method.
It is assumed that the fields $\phi_{\rm q} $, and $\phi_{\rm cl} $ are slow-varying ones, such that the Fourier-transformed fields have only long-wave components: $|k|<k_0$; $\omega <\omega_0$.
At the first step of RG transformations one integrates the partition function over the components of the fields in the limited wave band $\Lambda k_0<k<k_0$, $\Lambda^z \omega_0<\omega <\omega _0$. The renormalized parameters have the following form:
\begin{equation}\label{CH8R0}
\begin{array}{l}
   \displaystyle \mu^{(R)}=Z_{\mu}Z_{\phi_{\rm q }}Z_{\phi_{\rm cl} }\Lambda^{d+\varepsilon+z}=Z_{\mu}\Lambda^{-2},\\
   \displaystyle \gamma^{(R)}=Z_{\gamma}Z_{\phi_{\rm q} }^2\Lambda^{d+\varepsilon+2z}=Z_{\gamma}\Lambda^{0},\\
   \displaystyle v^{(R)}=Z_{v}Z_{\phi_{\rm q}}Z_{\phi_{\rm cl}}^3\Lambda^{3d+3\varepsilon+3z}=Z_{v}\Lambda^{\varepsilon}.
\end{array}
\end{equation}
where $Z_{\phi_{\rm q }}, Z_{\phi_{\rm cl} }, Z_{\mu}, Z_v$ and $Z_{\gamma}$ are the constants of renormalization.

Let us explain the renormalization of $\mu $ as an example in detail. We will limit ourselves to using the one loop approximation, that is quite enough for the demonstration of all the features of the theory. In this case the graphical representation of the main divergent contribution to the renormalization is shown in Fig.\,\ref{f3}b, and the renormalization constant of $\mu$ has the form:
\begin{equation*}\label{CH8R0}
\begin{array}{l}
  \displaystyle Z_{\mu }\simeq \mu -\dfr{6\mu v}{(2\pi)^{3}}\int\limits_{\Lambda^z \omega _0}^{\omega _0}\int\limits_{\Lambda k_0}^{k_0} G^K(k,\omega)G^R(k,\omega)d{\bf k}d\omega =\\[12pt]
  \displaystyle =\mu -\dfr{12\mu v\pi^2}{\gamma (2\pi)^3}\int\limits_{\Lambda k_0}^{k_0}\dfr{dk}{k}=\mu -\dfr{3\mu v}{2\gamma \pi}\ln(1/\Lambda).
\end{array}
\end{equation*}
One can see that the integral in this expression introduces a logarithmically divergent contribution to $\mu $ renormalization if the momentum dimension is $d[k]\equiv d=2$. In this case we get the following expression for the renormalized value of $\mu $:
\begin{equation}\label{CH8R}
\mu^{(R)}=e^{2\xi} Z_{\mu}\simeq e^{2\xi}\left[ \mu -\dfr 32\dfr{\mu v}{\gamma \pi}\,\xi\right],
\end{equation}
where $\xi=\ln(1/\Lambda )$ is the logarithmically divergent factor.
In the same way one can get other terms of the renormalized action:
\begin{equation}\label{CH8R1}
   \displaystyle v^{(R)}=e^{\varepsilon\xi } Z_{v}\simeq e^{\varepsilon\xi }\left[v-\dfr92\dfr{v^2}{\gamma \pi}\,\xi \right].
\end{equation}

The contribution to the renormalization of the kinetic coefficient $|\omega |\gamma $ is proportional to $|\omega |$:
\begin{gather*}\label{CH8R2}
\gamma^{(R)} =\gamma -\dfr{3v^2 16\pi^4}{(2\pi)^6\gamma^2} \ln{(1/\Lambda )}=\gamma -\dfr{3v^2}{4\pi^2\gamma ^2}\,\xi .
\end{gather*}

Hence, in the one-loop approximation the renormalization group of the model under study has the form:
\begin{equation}\label{RG1}
\begin{array}{c}
   \displaystyle \dfr{\partial \ln \mu}{\partial \xi}=2-\dfr 32\dfr{v}{\gamma \pi},\quad
   \displaystyle \dfr{\partial \gamma}{\partial \xi}= -\dfr 34\dfr{v^2}{\pi^2\gamma ^2},\\[10pt]
   \displaystyle \dfr{\partial \ln v}{\partial \xi}= \varepsilon -\dfr 92\dfr{v}{\gamma \pi}.
\end{array}
\end{equation}
 From the condition of the stable point existence, ${\partial \ln (v)}/{\partial \xi}=0$,  we obtain
$v=2\gamma\pi\varepsilon/9$.
Note that in the case of $d=2$ ($\varepsilon =0$) we get $v=0$. In this case only the quadratic term is relevant so that with $d=2=d_c^+$ the critical behavior is well described by the Gaussian theory.

From the above one can conclude, that there is the quantum order-disorder phase transition in one-dimensionalal systems. This differs appreciably from the classical case, in which the thermal fluctuations control the relaxation processes. However, this result agrees with the experimental data for the quasi-one-dimensional systems. Also, from (\ref{RG1}) one can see that in the one loop estimation the critical exponent $\nu =0.6$, and one can predict that the relaxation time, or equivalently, the phase coherence time, diverges at the critical point $g_c$ as $\tau \propto\gamma/\mu\propto -\ln|g-g_c|/(g-g_c)$.

\section{conclusions}

For Bose systems we have formulated the crossover from the critical dynamics regime of QPT to the quantum critical dynamics regime in the framework of the non-equilibrium quantum fields theory in the Keldysh technique. The key point of this crossover is that the random noise becomes pink when quantum fluctuations dominate over thermal fluctuations, $\hbar\omega >kT$. As a result, the system goes into a different class of universality, and the critical interval shifts into the low dimension area: $0<d<2$.

Experimental observations of QPT show that they can take place in one-dimensional systems. This conforms both with the QM approach and with the QCD description. However, according to QCD in the case of two-dimensional system at $T\to0$ the critical exponents should approach the mean-field theory values. This conforms with the recent experimental results \cite{Zhang}, and distinguishes this theory from QM approach, in which it should be possible only in the case of $d>3$.

One can assume that the reason of this is following: Indeed, at the time scales which are much more then the phase coherence time the description of the phase transition needs the QM approach. However, at the phase transition the phase coherence time diverges, and the experimental time scales can not exceed it. As a result the observed critical behaviour corresponds to the QCD description.

\begin{acknowledgements}
I am grateful to N.\,M.\,Shchelkachev and V.\,N.\,Ryzhov for helpful discussion of this paper.
This work was partly supported by the RFBR grants No. 13-02-91177 and No. 13-02-00579.
\end{acknowledgements}

\end{document}